# Categorize Interstellar Infrared Spectrum by Polycyclic Pure-Carbon-Molecule and Hydrocarbon-Molecule

Norio Ota

Graduate School of Pure and Applied Sciences, University of Tsukuba,
1-1-1 Tenoudai Tsukuba-city 305-8571, Japan;   n-otajitaku@nifty.com

By applying quantum chemical calculation, interstellar infrared spectrum (IR) was categorized to three classes. Type-A is unusual feature showing strong peaks at 11.3, 12.9, and 14.0 micrometer. Ubiquitously observed common major bands at 6.2, 7.7, and 8.6 micrometer were so weak or not recognized. Typical examples are NGC1316 and NGC4589, which include dark dust lanes originated from galaxy merger. Such unusual IR could be identified for the first time by pure carbon molecule $(C_{23})^{2+}$ (di-cation) having two carbon pentagons combined with five hexagons. Calculated spectrum coincided well with observed major bands at 11.3, 12.1, and 14.0 micrometer. Also, we could find other coincidence at 5.2, 5.6, 7.6, 8.8, 10.6, 15.7, and 17.2 micrometer. Pure carbon molecules play important part in interstellar IR. Type-B is ubiquitously observed IR, but showing medium strength at 11.3 micrometer compared with 6.2 micrometer one. Examples are NGC6946 and the red triangle nebula, which include active star burst region. Previously predicted polycyclic hydrocarbon $(C_{23}H_{12})^{2+}$ show best coincidence with observed many peaks. Also, spectrum intensity was reproduced well. Type-C is usually observed one featuring very strong peak at 11.3 micrometer. Examples are NGC7023, NGC2023 and M17SW, mostly in Milky Way and in matured giant galaxies. One capable explanation of large 11.3 micrometer band is a mixture of type-A and type-B. Another explanation is a mixture of type-B and none ionized PAH as like a neutral coronene $(C_{24}H_{12})$. Mixing degree of molecule species may vary 11.3 micrometer band strength. Combination of polycyclic pure-carbon-molecule and hydrocarbon-molecule could give good explanation on a variety of observed IR spectrum.

## 1, INTRODUCTION

Interstellar infrared spectrum (IR) due to polycyclic aromatic hydrocarbon (PAH) was ubiquitously observed in our Milky Way and other many galaxies from 3 to 20μm (Boersma et al. 2013, 2014). However, any single PAH molecule or related species showing universal infrared spectrum has not yet been identified to date. Identification is essentially important to search chemical evolution step of organics and to study material building block of life in the universe (Ota 2016). In 2014, accompanying a material study of void induced graphene sheet (Ota 2014a), it was founded for the first time that PAH molecule $(C_{23}H_{12})^{2+}$ shows very similar infrared spectrum with interstellar observed one (Ota 2014b, 2015a). This molecule contains two hydrocarbon pentagons combined with five hexagons. This may lead to an identification of specific carrier molecule for interstellar infrared emission. After that, many PAH molecules were test. Simple molecule $(C_{12}H_8)^{3+}$ had also show good coincidence with observed strong bands, which configuration was hydrocarbon one pentagon combined with two hexagons (Ota 2015b).  Recently, based on observation of nebula NGC 2023 (Peeters 2017), we could find out that among 16 observed bands, 14 bands were successfully reproduced well by calculation (Ota 2017b). Also, we could confirm such reproducibility even on galaxy scale observation (Ota 2017c). In order to understand such coincidence, I traced a history of star's death and birth and made one possible hypothesis named "Top down interstellar chemical evolution model" (Ota 2017a), that is, nucleation of graphene molecule during supernova expansion, void creation by high speed proton, hydrogenation by low speed proton, and ionization by high energy photon.

In this paper, such "Top down model" will be extended to pure carbon polycyclic molecules and observed IR will be categorized to three classes as like Type-A spectrum carried by pure carbon molecule, Type-B by hydrocarbon pentagon-hexagon combined molecule, and Type-C by a mixing of Type-A and Type-B.

2, TOP DOWN CHEMICAL EVOLUTION MODEL

In our previous study (Ota 2017a), we discussed "Top down hydrocarbon chemical synthesis model in the universe". Here, we like to add pure carbon molecule as shown in Figure 1. Interstellar dust was assumed to be created by an explosion of old star, for example, supernova (Nozawa 2003, 2006). After explosion, graphene molecule would be ejected to surrounding space, which may become a seed for organic dust. Typical model is ($C_{24}$) having pure carbon seven hexagons (coronene skeleton) as shown in (a). Ejected carbon diffuses to surrounding space at a high speed and will collide with interstellar gas, mainly with proton $H^+$. As shown in (b), high speed collision will make a void inside of graphene. It is essential that there occurs quantum mechanical configuration change as shown in (c), where void graphene will be rearranged to ($C_{23}$) having two carbon pentagons combined with five hexagons. Such polycyclic carbon molecule may cruise interstellar space long-long time. At a chance of new star born situation, such carbon seed will be illuminated by high energy photon. There occurs photoionization as shown in (d), which brings ionized molecule as ($C_{23}$)$^{n+}$. Ionization degree "n" depends on the central star radiated photon energy. Such molecule will bring Type-A infrared spectrum as discussed in a following chapter. Another situation will bring Type-B spectrum. At usual cases of new star born, there accompanies low speed proton irradiation (solar wind) to surrounding space, which may cause hydrogenation of graphene molecule as noted in (e). At the same time, there occurs photoionization (f) resulting cation PAH as like ($C_{23}H_{12}$)$^{2+}$.

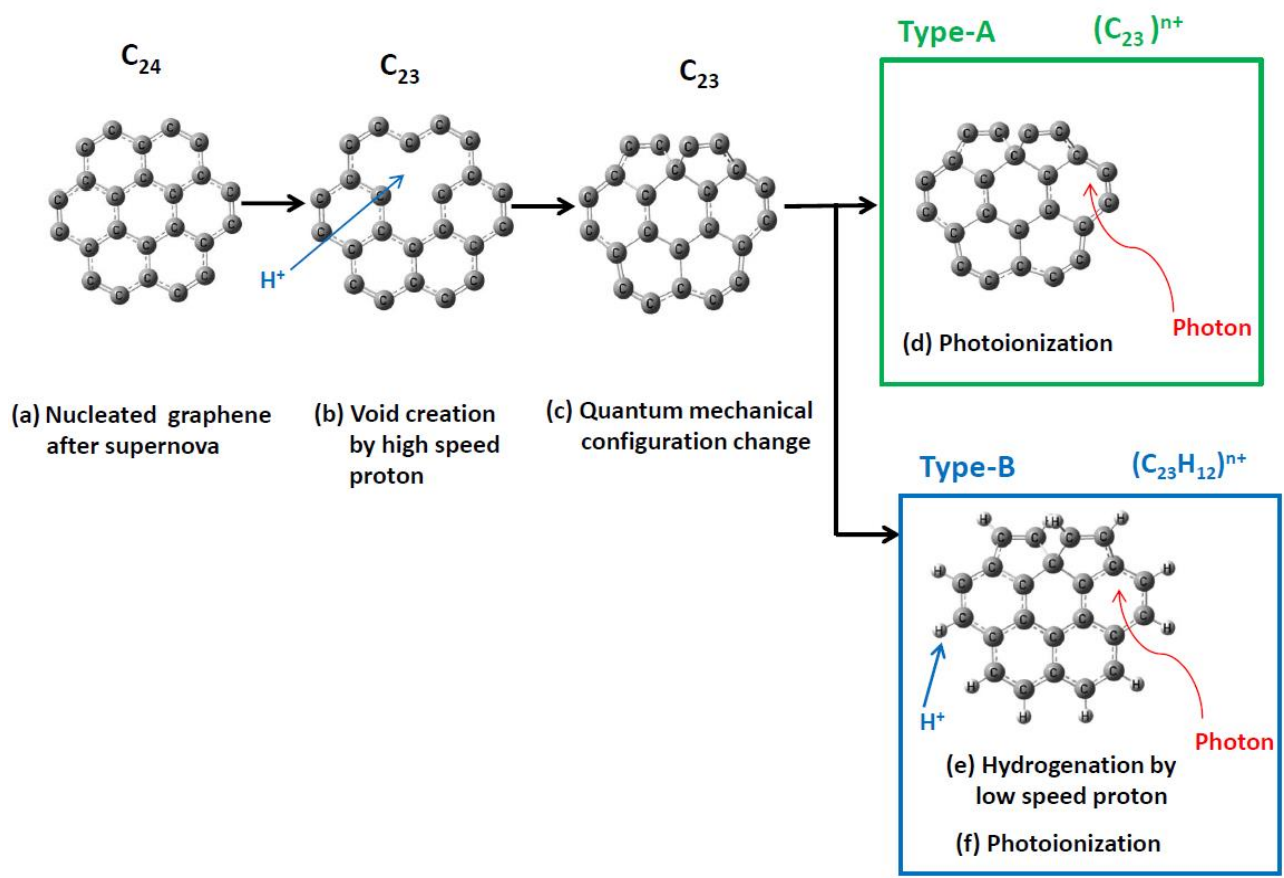

Figure 1, Top down interstellar chemical evolution model. Starting from pure carbon polycyclic molecule, by proton attack and photon irradiation, molecules would be modified to ionized polycyclic pure carbon (Type-A) and to ionized polycyclic hydrocarbon (Type-B).

## 3, CALCULATION METHOD

In quantum chemistry calculation, we have to obtain total energy, optimized atom configuration, and infrared vibrational mode frequency and strength depend on a given initial atomic configuration, charge and spin state Sz. Density functional theory (DFT) with unrestricted B3LYP functional was applied utilizing Gaussian09 package (Frisch et al. 1984, 2009) employing an atomic orbital 6-31G basis set. The first step calculation is to obtain the self-consistent energy, optimized atomic configuration and spin density. Required convergence on the root mean square density matrix was less than $10^{-8}$ within 128 cycles. Based on such optimized results, harmonic vibrational frequency and strength was calculated. Vibration strength is obtained as molar absorption coefficient ε (km/mol.). Comparing DFT harmonic wavenumber $N_{DFT}$ (cm$^{-1}$) with experimental data, a single scale factor 0.965 was used (Ota 2015b). Concerning a redshift for the anharmonic correction, in this paper we did not apply any correction to avoid over estimation in a wide wavelength representation from 2 to 30 micrometer.

Corrected wave number N is obtained simply by N (cm$^{-1}$) = $N_{DFT}$ (cm$^{-1}$) x 0.965.

Wavelength λ is obtained by λ (micrometer) = 10000/N(cm$^{-1}$).

Reproduced IR spectrum was illustrated in a figure by a decomposed Gaussian profile with full width at half maximum FWHM=4cm$^{-1}$.

## 4, TYPE-A INFRARED SPECTRUM

Typical examples of Type-A infrared spectrum were illustrated in Figure 2 from observations of NGC1316 reported by B. Asabere et al. (Asabere 2016) and NGC4589 by J. D. Bregman et al. (Bregman 2006). Type-A show strong peaks at 11.3, 12.9, and 14.0 micrometer, whereas common major bands at 6.2, 7.7, and 8.6 micrometer were so weak or not recognized. NGC1316 is 60 million light years (Myr) away elliptical galaxy, which is famous as very bright radio galaxy. This also includes dark dust lane, may be caused by eating several small galaxies. NGC4589 is 108 Myr away elliptical galaxy showing a dark dust disk rapidly rotating around the major axis, may be caused by galaxy merger. In 2005, supernova 2005cz was discovered in NGC4589, which is hydrogen poor, helium rich type-Ib supernova. These information suggested that candidate molecule carrying Type-A spectrum may be hydrogen poor, that is, pure carbon molecule. Applying model (d) in Figure 1, IR's of ionized molecules of $(C_{23})^{n+}$ were calculated. Results were illustrated in Figure 3. It was amazing that among four spectrum from neutral n=0 to n=3, di-cation case $(C_{23})^{2+}$ show very good coincidence with observed one. In Figure 2, calculated spectrum was compared with observed spectrum. Major observed bands at 11.3, 12.9, and 14.0 micrometer were reproduced very well at calculated 11.3, 13.0, and 14.0 micrometer. Longer bands at 15.7, and 17.2 micrometer in NGC1316 were also calculated to be 15.7, and 17.0 micrometer. It was amazing that even weaker band at 6.8, 7.6, 8.0, 8.7, 9.6, and 10.6 micrometer show excellent coincidence between observation and calculation. Remained problem was 5 micrometer range spectrum as noted by a red square in Figure 2. In calculation, we recognized very strong peaks at 5.2, and 5.6 micrometer, but no significant large spectrum in observations of Figure 2. Question is detector response of Spitzer/IRS as shown in top panel as W3 (Asabere 2016), which show very weak response window on 5 micrometer range. We should expect another infrared telescope observation. In 2007, H. Kaneda et al. had opened observed data by AKARI/IRC on the same NGC1316 as shown in Figure 4. Detector's windows are NG(2.5-5.0), SG1(5.5-8.3), and SG2(7.4-13.0). Observed nine band peaks were very well reproduced by our calculation as noticed by dotted vertical lines. Considering 4.5 micrometer deep signal gap, we can roughly estimate a supposed background curve by a blue dotted one. It was amazing that large peaks at 5.2 and 5.7 micrometer could be successfully reproduced, also other strength ratio between 11.3, 12.2, and 12.9 micrometer bands were well reproduced by calculation. Pure carbon polycyclic molecules, that is, modified graphene molecules may play very important role in active galaxies.

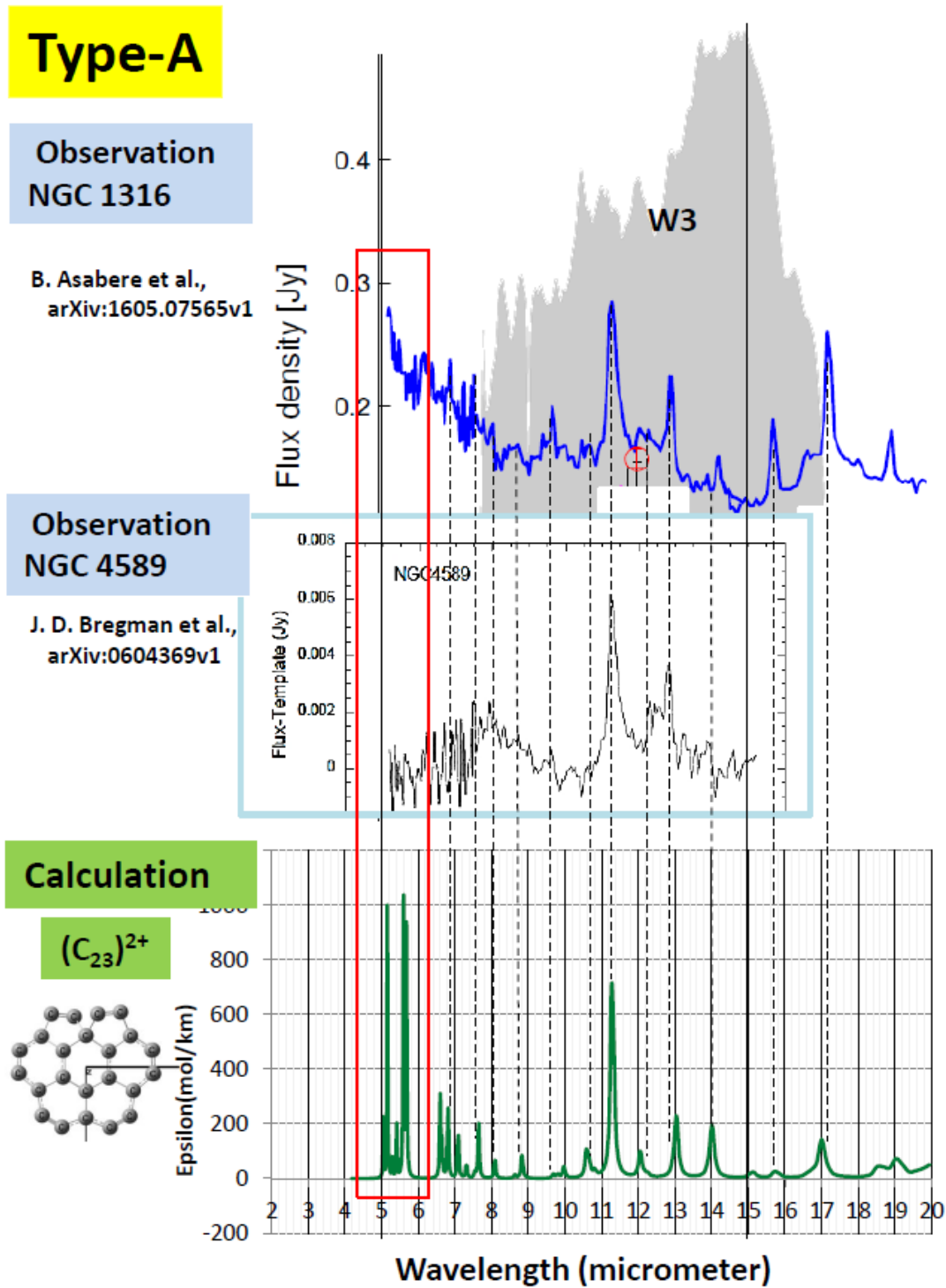

Figure 2, Type-A infrared spectrum. Typical examples are NGC1316 (Asabere 2016) and NGC4589 (Bregman 2006), which show strong peaks at 11.3, 12.9, and 14.0 micrometer, whereas ubiquitously observed PAH bands at 6.2, 7.7, and 8.6 micrometer were so weak or not recognized. Such unusual IR could be reproduced well by a pure carbon molecule $(C_{23})^{2+}$ having two carbon pentagons combined with five hexagons. However, as shown in a red box, calculated strong peaks at 5.2 and 5.6 micrometer are not clear in these two observations. Observed NGC1316 reported by Asabere may have a background increase in shorter wavelength out of detector window W3 (gray), while NGC4589 by Bregman show noisy response. This problem will be solved by another observation on NGC1316 by Kaneda (Kaneda 20007) as discussed in Figure 4.

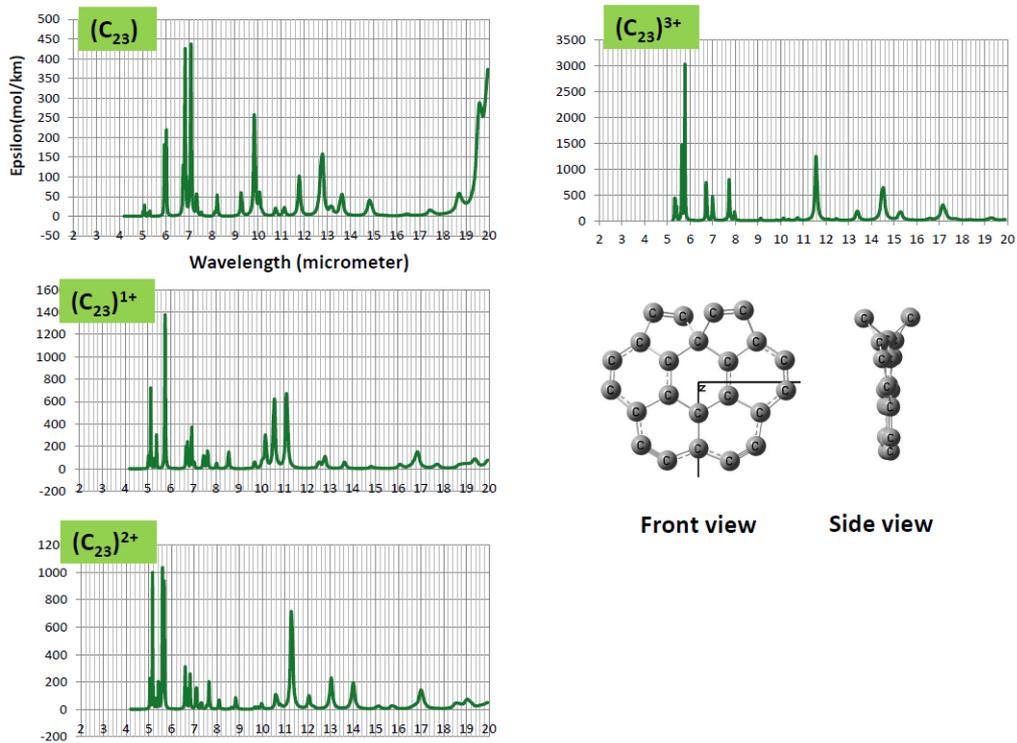

Figure 3, Calculated infrared spectrum on ionized polycyclic pure carbon molecule $(C_{23})^{n+}$ from n=0 to 3.

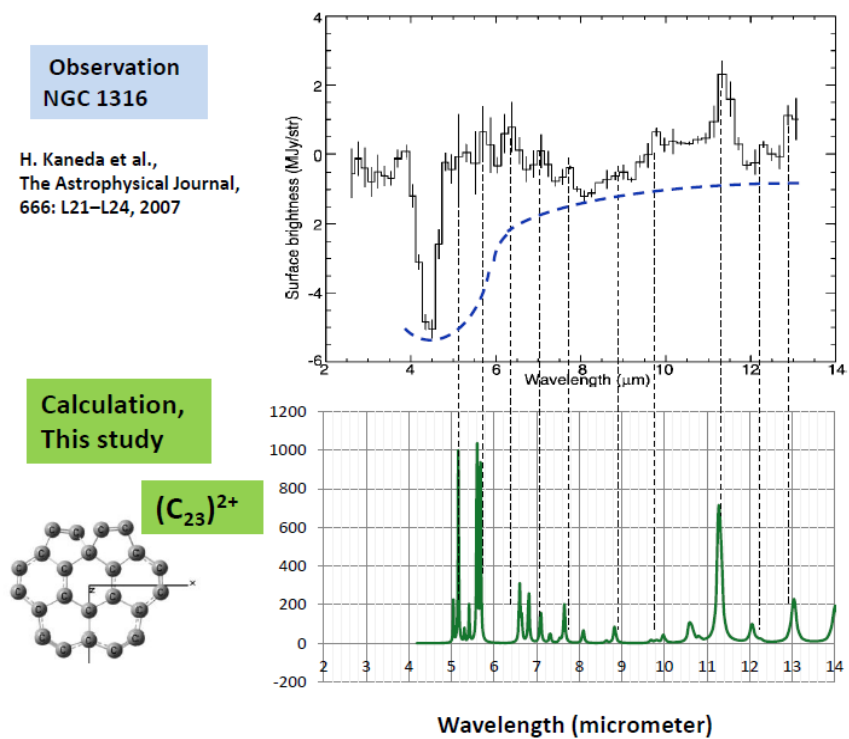

Figure 4, Wide range infrared observation by AKARI of NGC1316 (Kaneda 2007) on top panel. Very strong peaks at 5.2 and 5.6 micrometer in calculation of $(C_{23})^{2+}$ (down panel) could explain observed characteristics by assuming a blue dotted back-ground curve.

## 5, TYPE-B INFRARED SPECTRUM

Type-B infrared spectrum is illustrated in Figure 5, where examples are galaxy NGC6946 and the red triangle in Milky Way. Ubiquitously well observed 6.2, 7.7, 8.6, and 11.3 micrometer bands are observed as strong bands. Feature is medium strength of 11.3 micrometer band, where peak height is 40~80% comparing with that of 6.2 micrometer. Calculation result of polycyclic hydrocarbon $(C_{23}H_{12})^{2+}$ was shown in bottom of Figure 5, where major bands were well reproduced as 6.4, 7.7, and 8.5 micrometer. Calculated peak height at 11.2 micrometer was 80% comparing with that of 6.4 micrometer one. Both wavelength and strength were well reproduced. Observation of NGC6946 was done by I. Sakon (Sakon 2007) using AKARI/IRC. NGC6946 is 22Myr away, spiral galaxy including active star burst area as like an arm region. Mid panel show the red triangle nebula's IR spectrum using ISO-SWS data by G. Mulas et al. (Mulas 2006). Central star HD44179 would be reaching the end of its life and shedding its mass into space. Tiny peaks at 5.2, 5.6, and 9.6 micrometer in the red triangle (mid panel) may suggest some inclusion of Type-A in the red triangle dust cloud.

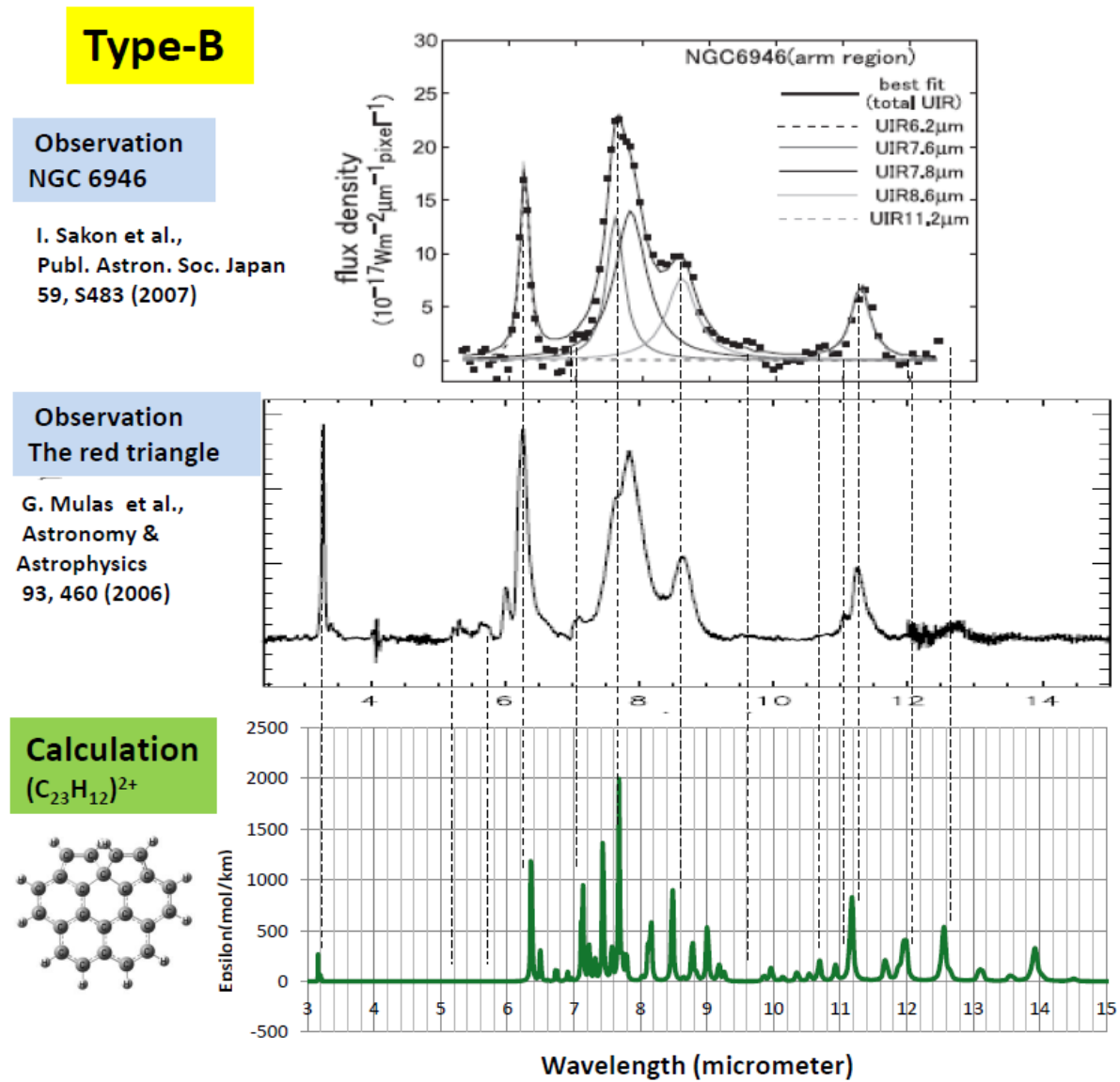

Figure 5, Type-B typical examples are NGC6946 galaxy arm region (Sakon 2007) and the red triangle nebula in Milky Way (Mulas 2006). Calculated spectrum of $(C_{23}H_{12})^{2+}$ could reproduce both wavelength and strength.

6, TYPE-C INFRARED SPECTRUM

Most common ubiquitously observed spectrum was categorized as Type-C. Observed examples are NGC7023 (Boersma 2013), NGC2023 (Peeters 2017) and M17SW (Yamagishi 2016) as shown in Figure 6. Major bands at 6.2, 7.7, 8.6 micrometer were observed well. Feature is very high peak at 11.3 micrometer than a height of 6.2 micrometer band. Also, 12.7 micrometer peak is relatively high than Type-B spectrum. It should be reminded that in Type-A, 11.3 and 12.9 micrometer bands are major strong bands. One idea of Type-C is a capability of a mixture of Type-A and Type-B. Figure 7 show calculated spectrum of $(C_{23}H_{12})^{2+}$ and $(C_{23})^{2+}$ comparing observed details of NGC2023. A red dotted vertical line show 11.3 micrometer position. If interstellar dust cloud would include those two carriers in some content, we can easily estimate an increase of 11.3 micrometer peak. However, there remains a problem that observed 5 micrometer bands are not major, weaker bands. Calculated 5 micrometer bands of $(C_{23})^{2+}$ are larger than 11.3 micrometer one. Another capability is a mixture of $(C_{23}H_{12})^{2+}$ and neutral PAH, as like $(C_{24}H_{12})$, which spectrum was illustrated on bottom in Figure 7. Problem of neutral PAH is that calculated strength of 3.3 micrometer band is similar height with 11.3 micrometer, whereas most observed 3.3 micrometer spectrum show weaker one. There remain unsolved questions on Type-C for both observation and theory.

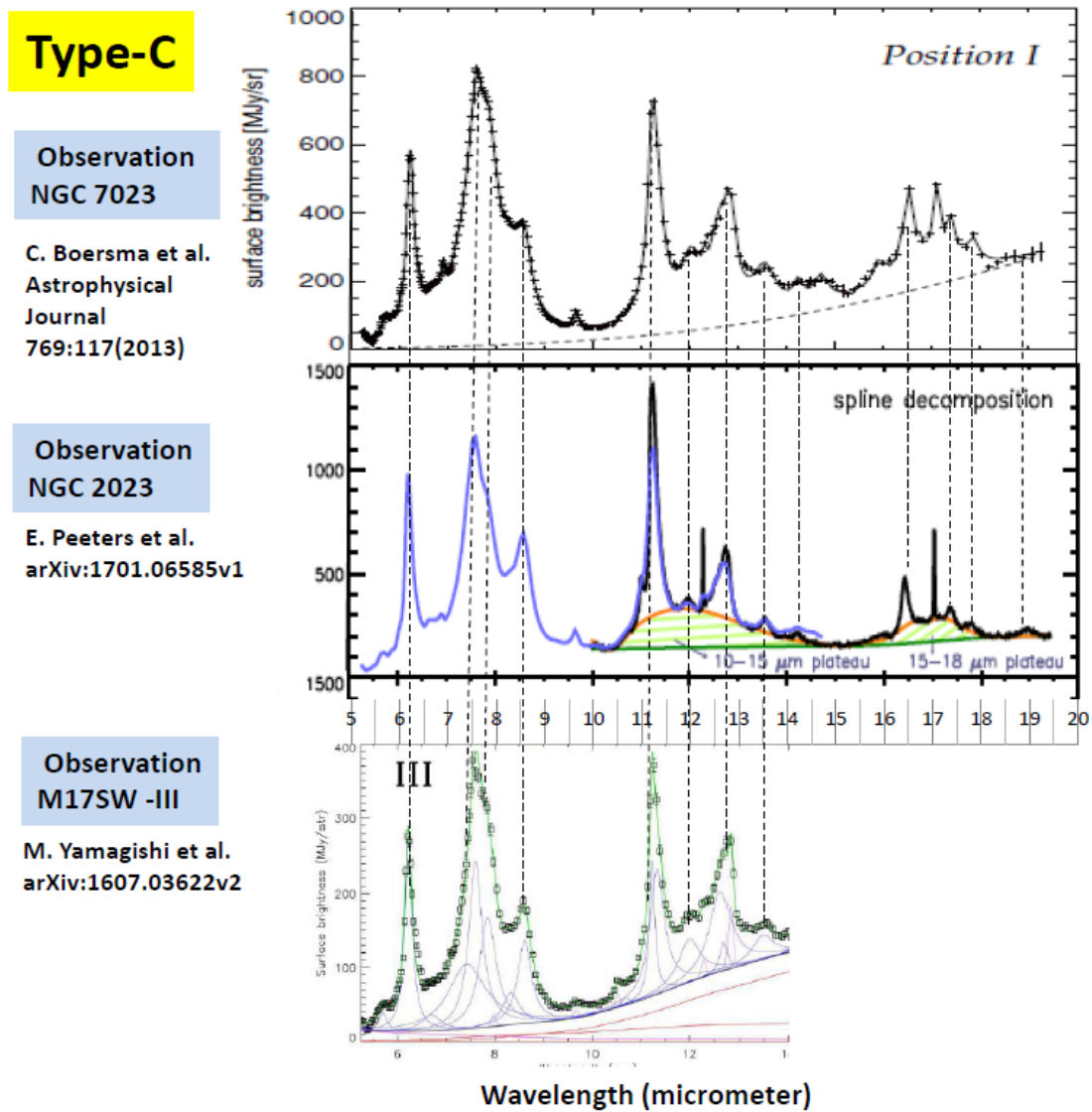

Figure 6, Type-C observation examples. Very strong peaks at 11.3 micrometer band are observed in NGC7023 (Boersma 2013), NGC2023 (Peeters 2023), and M17SW (Yamagishi 2016).

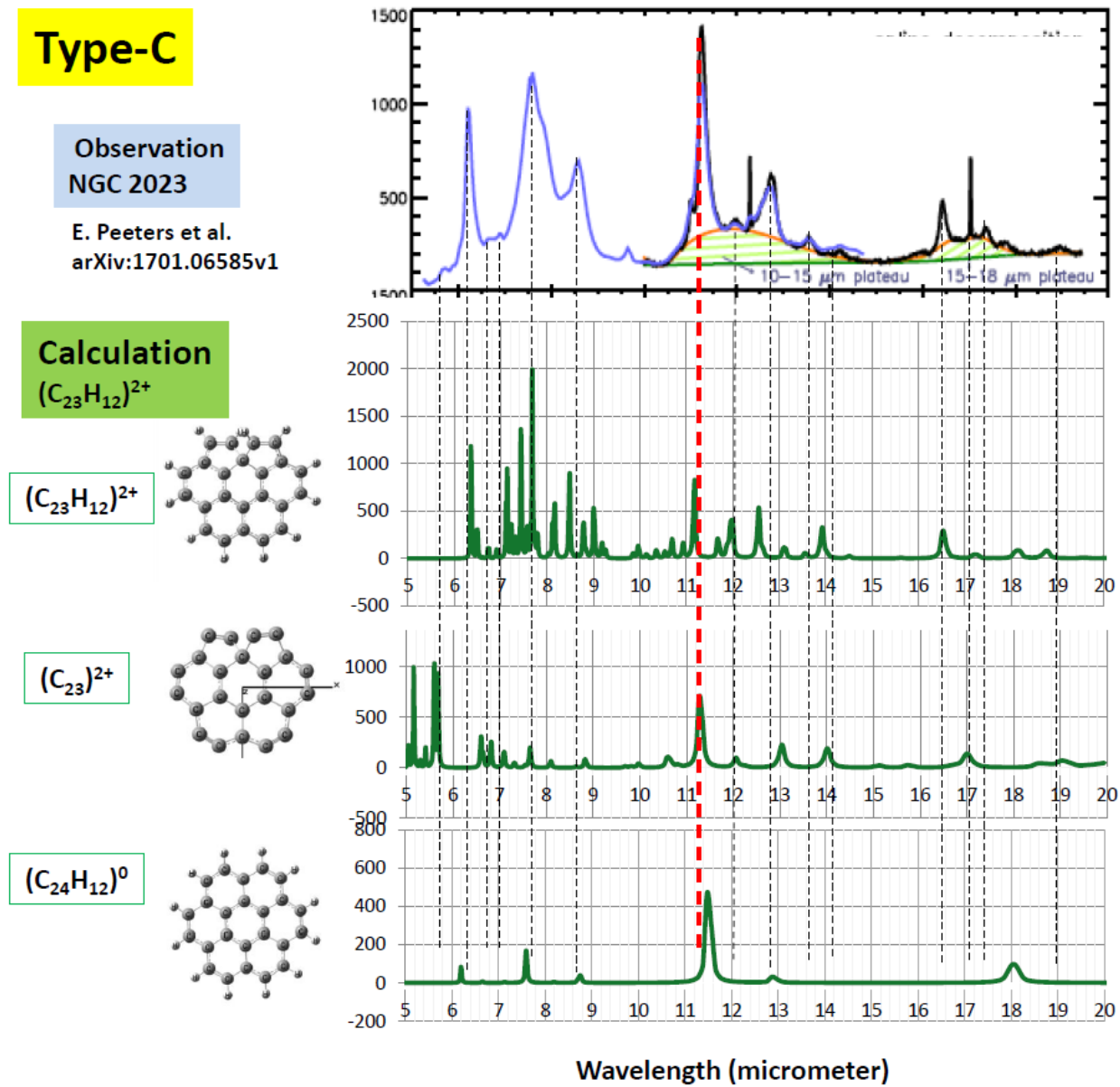

Figure 7, Trial of reproducing Type-C spectrum. One capability of reproducing strong 11.3 micrometer peak is a mixing of calculated $(C_{23}H_{12})^{2+}$ (Type-B) and $(C_{23})^{2+}$ (Type-A). Another capability is a mixing of calculated $(C_{23}H_{12})^{2+}$ (Type-B) and none ionized PAH as like $(C_{24}H_{12})$.

## 7, CONCLUSION

Interstellar polycyclic molecules would be carrier candidates of ubiquitously observed infrared spectrum (IR). Based on the top down interstellar chemical evolution model, observed spectra were categorized to three classes using quantum chemistry calculation.
(1) Type-A show strong peaks at 11.3, 12.9, and 14.0 micrometer, whereas common major bands at 6.2, 7.7, and 8.6 micrometer were so weak or not recognized. Typical examples are NGC1316 and NGC4589, which experience galaxy merger. Quantum chemical calculation on pure carbon molecule $(C_{23})^{2+}$ (di-cation) having two carbon pentagons combined with five hexagons could reproduce 11.3, 12.1, and 14.0 micrometer bands, and more detailed bands at 5.2, 5.6, 7.6, 8.8, 10.6, 15.7, and 17.2 micrometer.
(2) Type-B is ubiquitously observed IR, but showing medium strength at 11.3 micrometer band. Examples are NGC6946 and the red triangle nebula. Hydrocarbon pentagon-hexagon combined molecule $(C_{23}H_{12})^{2+}$ show best coincidence with observed many peaks. Also, every intensity ratio coincided well between observation and calculation.
(3) Type-C is ubiquitously most common observed spectrum, which showing very strong peak at 11.3 micrometer rather than case of Type-B. Examples are NGC7023, NGC2023 and M17SW, many in Milky Way and matured large galaxies. Several capabilities of such variation were considered as like a mixture of type-A and type-B, and also a mixture of type-B and none ionized PAH as like neutral coronene $(C_{24}H_{12})$. Mixture degree of dust may vary 11.3 micrometer band strength.

Both polycyclic pure carbon molecule (modified graphene) and modified PAH could give good explanation on a variety of observed IR spectrum.